\documentclass{vgtc}                          

\ifpdf
  \pdfoutput=1\relax                   
  \usepackage{graphicx}                
  \DeclareGraphicsExtensions{.pdf,.png,.jpg,.jpeg} 
\else
  \usepackage{graphicx}                
  \DeclareGraphicsExtensions{.eps}     
\fi%

\graphicspath{{figures/}{pictures/}{images/}{./}} 

\usepackage{microtype}                 
\PassOptionsToPackage{warn}{textcomp}  
\usepackage{textcomp}                  
\usepackage{mathptmx}                  
\usepackage{times}                     
\usepackage{cite}                      
\usepackage{tabu}                      
\usepackage{booktabs}                  

\vgtccategory{Research}

\vgtcinsertpkg

\title{Comparative Study of AR Versus Image and Video for Exercise Learning}

\author{Jamie Burns\\ %
    \scriptsize Birmingham City University  %
\and Wenge Xu\thanks{corresponding author, e-mail: wenge.xu@bcu.ac.uk}\\ %
     \scriptsize DMT lab \\\scriptsize Birmingham City University %
\and Ian Williams\\ %
     \scriptsize DMT lab \\ \scriptsize Birmingham City University
\and Irfan Khawaja\\ %
     \scriptsize School of Health Science\\\scriptsize Birmingham City University}

\abstract{There is inadequate attention to using mobile Augmented Reality (AR) in fitness, despite mobile AR being easy to use, requiring no extra cost, and can be a powerful learning tool. In this work, we present a mobile AR application that can help users learn exercises with a virtual personal trainer. We conduct a user study with 10 participants to investigate the learning quality of the ARFit (i.e., the proposed mobile AR application) in comparison to traditional methods such as Image-based learning and Video-based learning. Our results indicate that participants have a higher learning quality of exercise with mobile AR than (1) Image-based learning among all exercises selected and (2) video-based learning with exercise that requires greater spatial knowledge, with the performance evaluated by a qualified personal trainer. In addition, ARFit has an excellent rating in usability, is deemed to be highly acceptable, and is the preferred exercise learning method by most participants (N=9).
} 

\CCScatlist{
  \CCScatTwelve{Human-centered computing}{Human computer interaction (HCI)}{Interaction paradigms}{Mixed / augmented reality};
  \CCScatTwelve{Applied computing}{Education}{Interactive learning environments}{}
}

\begin{document}


\firstsection{Introduction}

\maketitle


Physical inactivity has been identified as one of the leading causes of non-communicable diseases such as cancer, diabetes, respiratory, and cardiovascular diseases \cite{lee_effect_2012}. It is estimated that over 1.6 million deaths per year can be attributed to physical inactivity \cite{GBD_global_2016}. There are many reasons why the general public is physically inactive, the literature suggests that possible factors could be a lack of intrinsic motivation, a lack of enjoyment, concern for perceived physical appearance, or social anxiety while exercising are contributing to physical inactivity among younger adults \cite{su13063183}. Additionally, barriers such as long transportation time/distance to the fitness centre, poor weather, and high membership fees have also prevented many adults from regular exercise \cite{ijerph18147341}.  

Many actions have been established to tackle the problem of physical inactivity among the general public, including virtual reality (VR) and augmented reality (AR). For instance, several VR exergames (e.g., FitXR\footnote{\url{https://fitxr.com/}}, VRworkout\footnote{\url{ https://github.com/mgschwan/VRWorkout}}, Dance Central\footnote{\url{https://www.dancecentral.com/}}) have been developed to engage the general public in exercising through exergaming. Smart mirror has been used to stream unlimited classes ranging from boxing to yoga, allowing users to exercise at home with a digital instructor, however, the cost of smart mirrors is expensive (e.g., Mirror \$1.5k, Tonal, \$3k, and ProForm Vue \$1.5k). Mostajeran et al. \cite{AR_elderly_training} presented a HoloLens-based AR coaching system for older adults to train their balance at home. The results suggest that older adults find the system encouraging and stimulating. The virtual coach is perceived as an alive, calm, intelligent, and friendly human. However, the cost of such a device (i.e., HoloLens) is not consumer-friendly (\$2k).

One approach that is consumer-friendly, highly acceptable, and allows users to exercise at home is using smartphone technology. Research suggests that the smartphone Health \& Fitness Market has already become popular with nearly 100k applications available on Google Play and about 160k applications on the Apple App Store in June 2021 \cite{yang_determinants_2021}. In this work, we aim to implement a mobile AR application (i.e., ARFit) for learning exercises with a virtual personal trainer, so that users can learn how to perform the exercise at home at any time without travelling, paying the membership, and concerning the poor weather. We then present a user study with 10 participants that compare the learning quality of using ARFit (AR-based learning) against traditional learning methods such as paper-based learning and video-based learning. Our results suggest that (1) participants have a higher exercise learning quality when using AR than (i) paper-based learning in all exercises and (ii) video-based learning in the exercise where spatial knowledge is needed, (2) learning exercises at home with the virtual personal trainer (ARFit) is preferred, feasible, and acceptable.

\section{Related Work}
We reviewed the literature on 1) AR and Fitness and 2) AR in Teaching and Learning.
\subsection{Augmented Reality and Fitness}
There are some attempts on integrating AR for fitness purposes, especially for therapeutic exercise. For instance, Luo et al. \cite{AR_finger_rehabilitation} implemented an AR system to render the stroke patient’s real hands inside a virtual environment for them to train their grasp-and-release ability by grasping and moving virtual objects. In addition, AR has been used in training balance \cite{AR_elderly_training}, hand skills \cite{AR_Hand_training}, and gait adaptability \cite{van_de_venis_improving_2021}.

Despite these successes in therapeutic exercise, there is still a very limited attempt to use AR to support, learn, or teach general exercise \cite{ng_effectiveness_2019}. Hsiao developed an AR learning system that combines exercise learning activities with regular academic lessons in physical education (PE). A 4-week study with 108 university students showed that (1) students make significant positive progress after using the AR application for 4 weeks (3 times a week) and (2) students have higher marks in the AR learning group than in the traditional teaching group (PowerPoint) \cite{hsiao_using_2013}. Interviews conducted with PE teachers indicate that they believe learning exercises in AR are not only more interesting and attractive, but also raise students' interests and motivations \cite{hsiao_using_2013}.

Similar results have been observed among seventh-grade students regarding motor skills learning \cite{motor_skill_AR}, where AR has been applied as a complement to video-assisted instruction, students can learn with interactive 3D models. Findings suggest that AR-assisted instruction is more effective than video-assisted instruction, and the effects are better for more difficult motor skills learning. Moreno-Gurrero et al. \cite{high_school_PE} have explored the impact of training activities through the use of AR in high school PR for the development and acquisition of spatial orientation, as opposed to the traditional exhibition training method. The results indicate that students (N=140) who participated in the AR group obtained greater motivation and grades (given by the teachers), especially in terms of spatial orientation. 

Based on this literature, it can be concluded that AR has shown the potential for fitness training and can be used in teaching and learning. However, the current attempts are mostly related to therapeutic exercises, general exercises such as Jumping Jack, Squat, and Sit Up should also be included. In addition, the current usage of AR in exercise learning and teaching is limited within the education context (school and university), there should be an AR application that allows the general population to learn exercise. Therefore, in this work, we aim to design and develop a mobile AR application for the general population to learn and practice exercise before exercise in a real-life scenario.  

\subsection{Augmented Reality in Teaching and Learning}
The research and development initiatives of AR technology in learning scenarios have increased rapidly over the last few years. Wagner et al. have compared mobile AR to Paper and PC in learning art history and suggested that the mobile AR method is the most enjoyable \cite{AR_in_education1}. Chiang et al. \cite{chiang_augmented_2014} have developed a collaborative mobile AR application for learning natural sciences, where the students (1) are guided to interact with various learning scenarios, and (2) can discuss and provide feedback among peers. They found that AR provided an exciting experience and students gained a high level of satisfaction and increased motivation and confidence in the tasks presented. Morillo et al. \cite{morillo_comparative_2020} have developed a mobile AR application to provide instructions for maintenance operations. Their user study suggests that AR is the most effective approach to substitute the typical paper-based instructions in consumer electronics. Moreover, In a review of 32 papers on the learning effects of AR vs non-AR applications \cite{radu_why_2012}, the author suggests that AR increases content understanding, motivation, and knowledge retention, when compared to traditional media approaches like paper or video media.

AR has also been used for learning and teaching in university education. Shelton and Hedley \cite{1106948} found that AR can positively affect undergraduate Geography students’ comprehension of Earth-Sun relationships. Akcayir et al. \cite{AKCAYIR2016334} suggested that using AR to provide additional virtual learning content (i.e., texts, audios, videos, animations, and simulations) explaining how to set up and conduct the experiment, as well as some of the underlying phenomena can significantly help achieve exam scores than the non-AR condition where only paper-based laboratory manuals are used. Thees et al. \cite{THEES2020106316} suggested that learning physics laboratory courses with AR led to a significantly lower cognitive load than the traditional workflow condition (i.e., PC).

Overall, AR has the potential to provide a positive (e.g., motivating, satisfying, engaging) and meaningful learning experience among all users \cite{nincarean_mobile_2013}. In this work, we aim to (1) compare our proposed mobile AR exercise learning application (i.e., ARFit) against traditional learning methods like Image and Video to test the learning effectiveness, and (2) evaluate the usability and acceptability of ARFit. 

\section{ARFit: Augmented Reality Exercise Learning Application}\label{ARFitsection}
This section describes details of the design and implementation of the proposed mobile AR exercise learning application (i.e., ARFit). The ARFit is used to augment a virtual personal trainer in the real world to perform a collection of animations based on the user's requirement. Below we describe how we implement the virtual personal trainer feature and the interaction feature to allow the user to interact with the virtual personal trainer. A video demonstration has been provided in the supplement video\footnote{\url{https://www.youtube.com/watch?v=3naFbmdCdf4}}. 

\subsection{Virtual Personal Trainer}
A vital part of the ARFit is to have a virtual personal trainer perform a variety of exercises. This is achieved by using Adobe Mixamo, which has a range of 3D models and animations to select from and has 3D model files specifically for Unity (i.e., in .FBX format). After importing the models into Unity, several modifications are required to make sure the textures (Unity setting: Location setting in Legacy and then change to Use Embedded Materials) and rigging (Unity setting: Animation type as Humanoid) are correctly set up.

\begin{figure}[h]
  \centering
  \includegraphics[width=\linewidth]{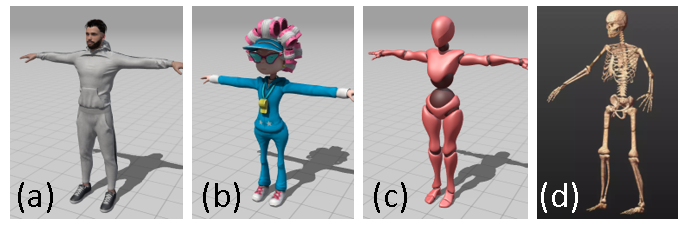}
  \caption{Four types of avatar tested in the pilot study: (a) realistic human, (b) cartoon, (c) humanoid, (d) skeleton.}
  \label{fig:Avatar}
\end{figure}

We conducted a short pilot study with 16 participants (11 male) to decide which avatar style (i.e., realistic human, cartoon human, humanoid, and skeleton; see Fig. \ref{fig:Avatar}) should be used in our experiment. Among these avatar styles, most (N=13) participants selected the realistic human avatar for the virtual personal trainer. The supporting reasons collected are that a realistic human avatar is more convincing (looks like a real personal trainer), while others argue that the realistic virtual personal trainer can help them identify which body part or muscle groups should move while other avatar styles could be misleading. Based on the positive feedback collected from this pilot study, we have decided to use the realistic human model as the default avatar of the virtual personal trainer in ARFit. 

\subsection{Interaction}
The ARFit is developed in Unity3D (v2020.3.27f1) with the Vuforia AR SDK\footnote{\url{ https://developer.vuforia.com/}}. Vuforia comes with a variety of prefabs of AR technologies to provide a starting point for an application, ARFit utilises the ARCamera, Ground Plane, Plane Finder, and Image Targets prefabs.

The ARCamera acts as the main viewing camera, allowing users to view the AR feature (i.e., the virtual personal trainer) in ARFit. The Plane Finder prefab together with the Ground Plane prefab allows users to track any horizontal surface from ground to smartphone to place a virtual personal trainer into the real environment to show users how to perform exercises. A reset ground plane function has also been added so that users can reset the plane if they made a mistake or find a better horizontal surface. 

One disadvantage of the Ground Plane solution is that users need to have clear and empty space in their real environment to place the virtual personal trainer. This may not be feasible for people who are living in limited spaces. To overcome this issue, we have employed the Image Target prefab, where users can use their smartphone to scan the QR code to view the augmentation of the linked 3D model on top of it (i.e., the virtual personal trainer with different exercise animations). This process makes use of Vuforia Library’s database to store the Image Targets (i.e., QR codes). 

Once visualized the virtual personal trainer, users have the following interaction control via touchscreen buttons: (1) play/pause the animation, (2) rotate the 3D model, (3) reset the ground plane, (4) back to the main menu. In addition, they can walk around with the phone to change the camera viewport.

\section{Experiment}
To explore the effectiveness of exercise learning, we compared the ARFit against traditional learning methods such as learning with images and videos. The usability and acceptability of the ARFit were also evaluated. 

\subsection{Experiment Design}
To avoid any bias between participants, the study was conducted using a within-subject design where each participant would interact with all the required conditions (AR, Video, Image). The order of the conditions was counterbalanced with a Latin square.

\textit{AR}. In the AR condition, the participants needed to identify a horizontal surface first, and then place the virtual personal trainer in the real environment. To gain a better understanding of how to perform the exercise, participants can rotate the virtual personal trainer via touch screen or walk around the virtual personal trainer to find their preferred viewing angles. They can also pause or restart the animation depending on their needs (see Section \ref{ARFitsection}).

\textit{Video}. To avoid confounding factors, videos used in the experiment were screen recordings of exercises on ARFit. Each exercise screen recording was recorded in a static position, the exercise shown in the recording presents a clear procedure on how to perform the exercise. The quality of the exercise in the video was approved by a qualified personal trainer. In this condition, participants were able to pause or restart the video depending on their needs. Refer to the supplemented video\footnote{\url{ https://www.youtube.com/shorts/kM2_S1pmD1I}} for an example of the Video condition we used in the experiment. 

\begin{figure}[h]
  \centering
  \includegraphics[width=\linewidth]{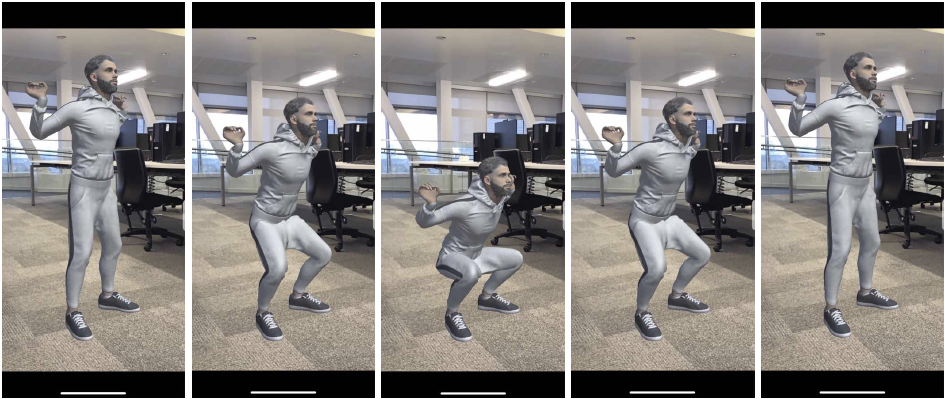}
  \caption{Example of Image learning method where the virtual personal trainer is performing a Squat. }
  \label{fig:Exercise}
\end{figure}

\textit{Image}. The step-by-step images used in this condition were captured based on the videos in the Video condition to avoid confounding factors. The step-by-step images covered all essential steps of how to perform the exercise and were evaluated and approved by a qualified personal trainer. Fig. \ref{fig:Exercise} shows an example of the step-by-step image instructions of the selected exercise used in this condition.

\subsection{Procedure, Tasks, and Apparatus}
Before the experiment began, participants needed to read and sign the information sheet and the consent form. Then, they were required to fill out a pre-experiment questionnaire that gathered demographic information (e.g., age, gender, knowledge of exercise). 

Before each condition, they were given instructions on how to use the corresponding learning method. For each condition (AR, Video, Paper; the order counterbalanced), participants were required to perform 4 types of exercises (Squat, Sit Up, Jumping Jack, and Burpee; in random order). For each exercise, they had one minute to view and remember it, and then, they needed to perform three repetitions of the corresponding exercise they just watched. Once they performed these repetitions, they proceeded to the next exercise until all these exercises were completed. Participants were given as much time as needed to perform the repetitions and they were allowed to rest in between exercises. 

Participants needed to fill out a system usability questionnaire after the AR condition. After completing all conditions, they were required to complete a post-experiment questionnaire (ranking and acceptability) and participate in a short structured interview to share their thoughts and provide feedback regarding the study and the ARFit. 

Each experiment lasted approximately 30 minutes and was conducted in safe environments. There was no compensation given to the participants. We used three devices in this experiment, a laptop for reading and signing the information sheet and consent form, filling out the questionnaire, an iPhone 11 for learning exercises, and an additional phone for recording participants' exercise performance. 

\subsection{Outcome Measure}
\textit{Learning quality}. The learning quality of each exercise was assessed by a qualified personal trainer. The personal trainer was asked to watch the recorded video of a participant performing exercises (e.g., Squat, Sit up, Jumping Jack, and Burpee) and then gave a score from 1 (Extremely Bad) to 10 (Extremely Good) based on the quality of the exercise. We employed a blind review process, where the personal trainer did not know which learning condition was used when rating the recording video.

\textit{Usability}. System Usability Scale (SUS) was used to measure the usability of the ARFit. SUS score ranges from 0 to 100, where a score of 100 represents the best usability and a score above 68 is considered positive \cite{sus_1996}. This questionnaire was used after the participants completed the AR condition. 

\textit{Acceptability}. We assessed the acceptability of the ARFit using a 5 point-Likert scale (1 indicated “Strongly Disagree” and 5 was “Strongly Agree”) question “How likely are you to recommend this app to a friend?” \cite{iVR_mental}. 

\textit{Ranking and Structured Interview}. Participants needed to select their preferred exercise learning method among AR, Paper, and Video. Once selected their preferred learning method, we conducted a short structured interview with three open-ended questions, “What was your reason for your preferred learning method selection?”, "Was there anything you didn’t like about the mobile AR application?”, “Have you performed all the exercises before?”. Answers were recorded and transcribed in text.

\subsection{Participants}
We gathered data from 10 participants (5 male), between 20 and 26 years of age (Mean = 22.1, SD = 1.60), recruited from our local university. 6 of the participants had used an AR application before but not for exercise. Only 2 participants described themselves as advanced in terms of their exercise knowledge. These 2 participants were also regularly exercised. 

\section{Results}
\subsection{Statistical Analysis}
IBM SPSS 28 for Windows was used for analysing the data. We used Shapiro-Wilks tests and Q-Q plots to check if the data had a normal distribution. All tests are with two-tailed p values. For normally distributed data, we employed One-way repeated-measure ANOVA with Learning Methods (AR, Video, Image) as the within-subject factor. Greenhouse-Geisser adjustment was used to correct violations of the sphericity assumption. For data that were not normally distributed, a Friedman Test was used and Wilcoxon Signed Rank Test was used for post-hoc analysis. Bonferroni corrections were used for all pairwise comparisons.

\subsection{Learning Quality}
\begin{figure}[h]
  \centering
  \includegraphics[width=\linewidth]{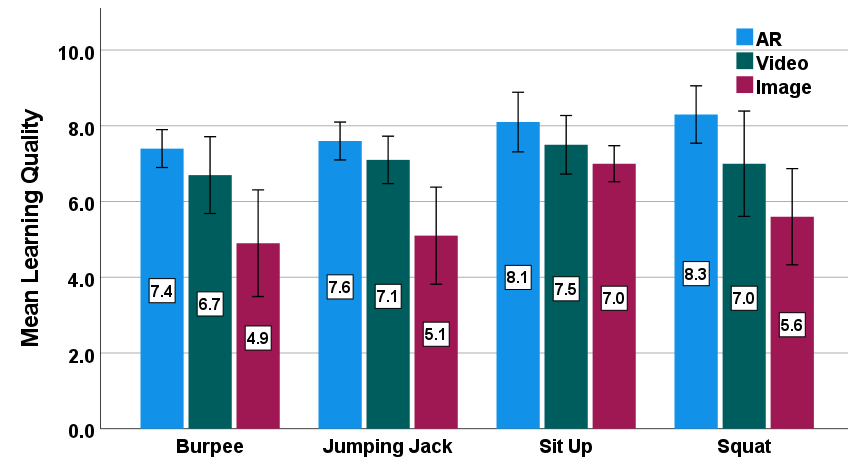}
  \caption{Mean learning quality score rated by a qualified personal trainer for Burpee, Jumping Jack, Sit Up, and Squat among three learning methods. Error bars indicate a 95\% confidence interval.}
  \label{fig:learningquality}
\end{figure}

Details of learning quality for each exercise can be found in Fig. \ref{fig:learningquality}. 

\textit{Squat}. One-way repeated measure ANOVA determined a statistically significant difference between the Learning Method on Squat learning quality ($F_{1.986,17.875} = 24.251, p < 0.001$). Post-hoc pairwise comparison (Bonferroni corrected) suggested learning quality of squat in AR was found to be significantly higher than in Image ($p < 0.001$) and Video ($p = 0.028$) conditions.

\textit{Jumping Jack}. Friedman test determined a statistically significant difference between the Learning Methods on Jumping Jack learning quality ($\chi^2(2) = 16.800, p < 0.001$). Post-hoc pairwise comparisons revealed that there was a significant difference between AR and Image ($Z = -2.692, p = 0.007$) and between Video and Image ($Z = -2.687, p = 0.007$) on Jumping Jack learning quality, we could not find a significant difference between AR and Video.

\textit{Sit Up}. Friedman test yielded a statistically significant effect of the Learning Method on Sit Up Learning Quality ($\chi^2(2) = 13.556, p = 0.001$). Post-hoc pairwise comparisons indicated that there was a significant difference between AR and Image ($Z = -2.810, p = 0.005$) on Sit Up learning quality, but no other significant differences were found.

\textit{Burpee}. Friedman test determined a statistically significant difference between the Learning Methods on Burpee learning quality ($\chi^2(2) = 15.793, p < 0.001$). The post-hoc analysis yielded that there was a significant difference between AR and Image ($Z = -2.692, p = 0.007$) and Video and Image ($Z = -2.585, p = 0.010$) on Burpee learning quality, we could not find a significant difference between AR and Video.

\subsection{SUS and Acceptability} 
The participant’s mean SUS score was 81.3, which suggests ARFit has an excellent score regarding usability (score of 68: okay; scores of 68$-$80.3: good; scores of $>$80.3: excellent). of the 10 participants, 6 participants rated it as excellent, another 3 rated it as good, and 1 rated it as poor. The lowest SUS score was 65 and the highest was 90. 

In addition, all of our participants agreed (N=2) or strongly agreed (N=8) that they would love to recommend this app to their friends.

\subsection{Ranking and Feedback}
Most participants (N=9) selected AR as their preferred method of exercise learning, only 1 participants selected Video as the preferred exercise learning method.

Regarding the reasons for selecting AR as their preferred method, participants provided a range of positive feedback, for instance, easy to use, clear, more interactive. Below shows selected comments from participants:
\begin{itemize}
  \item P1: “Even though the quality of images were clear, it was hard to understand (how to perform) the correct movements. Instantly with video, it was a lot easier but with (only) one angle it was different to see the specific movements and how body parts are set up like feet for example.”
  \item P4: “I found AR a lot more useful because I could see all the angles and move closer to the character to understand the form which helped me realise, I needed to keep my feet on the floor. It also made me realise I was doing the motion too quickly as the AR model was clear on the correct speed to do the exercise.”
  \item P10: “3D model allowed me to have a full view of exercise demonstrations allowing me to properly view form.”
\end{itemize}

Despite choosing Video as the preferred learning method, P9 did not dislike the AR application and had no issue with it.

All participants mentioned they had no problem with the AR application. They also suggest that the application could have more exercise choices.

Regarding whether they have performed all these exercises before, 8 participants mentioned they had performed these exercises before and 2 participants were new to them. Despite having experience performing these exercises, participants found the AR version helped them correct the movement (P2: ``(I) have done the exercises before but I did make changes to my form using the AR version."). Those who have never performed also agreed that the AR version was useful (P4: ``(I) have not performed any of the exercises before and this application was useful to help me understand the correct form and speeds needed when doing the exercises in an easy to use and fun application. I am sure it would help me understand many more exercises.").

\section{Discussion and Future Work}
Our results indicate that users learned better with AR than the Image-based method among all selected exercises, which is in line with previous literature on learning in physical education \cite{hsiao_using_2013} and other contexts such as manual assembly tasks \cite{8613759}, laboratory skills \cite{AKCAYIR2016334}, and navigation skills \cite{AR_navigation_skill}. Among traditional learning methods, the Video-based learning was better than the Image-based learning regarding most exercises (i.e., Squat, Jumping Jack, and Burpee), which is in line with the literature (e.g., health \cite{video_paper}). This suggests that videos can be provided as a useful exercise learning method when AR is not available.

We also observed that users learned better with AR than with Video in exercises like the Squat, one possible explanation is that users need to rotate the model to better view the details of the exercise spatially, which supports \cite{high_school_PE} that AR is showing spatial knowledge.

The usability of the app proved excellent from the SUS score of 81.3 suggesting that the application overall provided an easy-to-use experience for the user and this was backed by multiple responses in the interview feedback (e.g., “an easy to use and fun application”). AR exercise learning app not only took the lead on scores of learning quality, it was also the preferred learning method by most participants (9 out of 10).

A list of updates has been made in the latest iteration of ARFit based on the suggestions provided by participants. For instance, more exercises have been added, functions such as re-sizing the virtual personal trainer, changing the avatar of the personal virtual trainer (e.g., different body sizes), playing exercise animation at a variety of speeds (slow, normal, fast) to help better understanding, filtering exercise by body parts and difficulty level. An updated version of ARFit can be found in\footnote{\url{ https://www.youtube.com/watch?v=3naFbmdCdf4}}. 

Currently, the presented work only included university students as our participants, we have started working with wheelchair users and sports scientists to evaluate and modify the ARFit to make it more inclusive. A further study with wheelchair users will be conducted to investigate the feasibility, acceptability, and effectiveness of using ARFit for learning exercises. Moreover, the current study focused on plane scanning function and did not evaluate the QR code scanning function, this will be evaluated as part of our future work.

\section{Conclusion}
Barriers such as long time/distance to the fitness centre, poor weather, and high membership fees are making it difficult for people to learn exercises and become physically active. In this research, we have developed ARFit, a mobile AR fitness application for users (both physically active and inactive) to learn exercises with a virtual personal trainer. We have explored the learning quality of ARFit against traditional learning methods (i.e., Image and Video) and found AR outperformed Image in learning all selected exercises and outperformed Video in exercises where greater spatial information is desired. Our results also suggest that the ARFit has an excellent usability scale and is highly acceptable by users.

\acknowledgments{
We want to thank all the participants for their time and reviewers for their advice that helped improve our paper.}

\bibliographystyle{abbrv-doi}
\bibliography{IEEE.bib}
\end{document}